\input ./style/arxiv-general.cfg
\documentclass[aoas,MSNbibl,nameyear,dvips]{arximspdf}
\makeatletter
   \@ifpackageloaded{graphicx}{}{\usepackage{graphicx}}
\makeatother
\usepackage{multirow}

\doi{10.1214/15-AOAS813}
\volume{9}
\issue{2}
\pubyear{2015}
\firstpage{866}
\lastpage{882}
\docsubty{FLA}

\makeatletter
\newproclaim{defin}{Definition}
\makeatother

\begin{document}
\begin{frontmatter}

\title{Multi-species distribution modeling using penalized mixture of
regressions}
\runtitle{Multi-species modeling using mixtures}

\begin{aug}
\author[A]{\fnms{Francis K. C.}~\snm{Hui}\corref{}\thanksref{M1,T1}\ead[label=e1]{fhui28@gmail.com}},
\author[A]{\fnms{David I.}~\snm{Warton}\thanksref{M1,T2}\ead[label=e2]{david.warton@unsw.edu.au}}
\and
\author[B]{\fnms{Scott~D.}~\snm{Foster}\thanksref{M2,T3}\ead[label=e3]{scott.foster@csiro.au}}
\runauthor{F. K. C. Hui, D. I. Warton and S. D. Foster}
\affiliation{University of New South Wales\thanksmark{M1} and CSIRO\thanksmark{M2}}
\address[A]{F. K. C. Hui\\
D. I. Warton\\
School of Mathematics and Statistics\\
University of New South Wales\\
Sydney 2052\\
Australia\\
\printead{e1}\\
\phantom{E-mail:\ }\printead*{e2}}
\address[B]{S. D. Foster\\
CSIRO\\
Hobart 7001\\
Australia\\
\printead{e3}}
\end{aug}
\thankstext{T1}{Supported in part by a Research Excellence Award from
the University of New South Wales and a CSIRO PhD Scholarship.}
\thankstext{T2}{Supported in part by Australian Research Council
Discovery Projects and Future Fellow funding
schemes (project number DP130102131 and FT120100501).}
\thankstext{T3}{Supported in part by the Marine Biodiversity Hub, a
collaborative partnership
supported through funding from the Australian Government's National
Environmental Research Program (NERP).}

%
\received{\smonth{8} \syear{2014}}
%
\revised{\smonth{2} \syear{2015}}

\begin{abstract}
Multi-species distribution modeling, which relates the occurrence of
multiple species to environmental variables, is an important tool used
by ecologists for both predicting the distribution of species in a
community and identifying the important variables driving species
co-occurrences. Recently,
Dunstan, Foster and Darnell
[\textit{Ecol. Model.} \textbf{222} (2011)  955--963]
proposed using finite
mixture of regression (FMR) models for multi-species distribution
modeling, where species are clustered based on their environmental
response to form a small number of ``archetypal responses.'' As an
illustrative example, they applied their mixture model approach to a
presence--absence data set of 200 marine organisms, collected along the
Great Barrier Reef in Australia. Little attention, however, was given
to the problem of model selection---since the archetypes (mixture
components) may depend on different but likely overlapping sets of
covariates, a method is needed for performing variable selection on all
components simultaneously. In this article, we consider
using penalized likelihood functions for variable selection in FMR
models. We propose two penalties which exploit the grouped structure of
the covariates, that is, each covariate is represented by a group of
coefficients, one for each component. This leads to an attractive form
of shrinkage that allows a covariate to be removed from all components
simultaneously. Both penalties are shown to possess specific forms of
variable selection consistency, with simulations indicating they
outperform other methods which do not take into account the grouped
structure. When applied to the Great Barrier Reef data set, penalized
FMR models offer more insight into the important variables driving
species co-occurrence in the marine community (compared to previous
results where no model selection was conducted), while offering a
computationally stable method of modeling complex species--environment
relationships (through regularization).
\end{abstract}

\end{frontmatter}

\section{Introduction} \label{secintro}

Multi-species distribution modeling, which relates the occurrence of
multiple species to environmental covariates, is an important tool both
for predicting how a species community will respond to changing
environmental conditions and for identifying important environmental
variables driving species co-occurrences [\citet{ferrier06,ovaskainen2010modeling,pollock14}]. To construct such
models, statistical methods are required which can handle the
underlying heterogeneity in species--environment relationships (i.e.,
different species in the same community can have very different
environmental responses), while providing accurate predictions for rare
species that may not be modeled reliably on their own [by borrowing
strength across organisms in a community, \citet{ferrier06,ovaskainen2011making}].

Recently, \citet{dunstan11} proposed using finite mixture of regression
[FMR, \citet{wedel95}] models for multi-species distribution modeling
of presence--absence (binary) data, where spe\-cies are clustered based on
their environment response to form a small number of ``archetypal
responses.'' The methodology was extended by \citet{hui13} and \citet
{dunstan13} to handle count and biomass data, the latter being a
nonnegative continuous value representing the combined weight of all
individuals of each species. By clustering species into archetypes, and
modeling each archetype using a generalized linear model [GLM, \citet
{mccullagh89}], these Species Archetype Models or SAMs offer a powerful
approach to modeling heterogeneity in a community's response to a set
of covariates. Clustering species based on environmental response is
also consistent with recent findings in ecology that groups of species
tend to respond in a similar manner to environmental gradients [at
least with the resolution of most multi-species data sets, \citet{clark2010}]. Moreover, \citet{hui13} showed SAMs offer strong
predictive performance of rare species by borrowing strength from more
prevalent species classified to the same archetype.

While these initial results for SAMs showed promise, little attention
was given to the important issue of model selection. In their
application of SAMs to a data set of presence--absence records collected
for 200 species along the Great Barrier Reef off the northeast coast of
Australia, \citet{dunstan11} used a heuristic version of BIC [\citet
{schwarz78}] to select the ``types of covariates'' to enter in the
model, that is, physical habitat covariates, oceanographic measures or
both. Also, both \citet{hui13} and \citet{dunstan13} a priori
fixed the set of covariates to enter into their respective SAMs. In all
three articles, the \emph{same} set of covariates were entered into
each archetype. This is a restrictive requirement, as it fails to
account for the numerous (and unknown) ways which organisms react to
their environment.

Since the component densities may depend on different but likely
overlapping sets of covariates, a method is needed for performing
variable selection on all components in an FMR model simultaneously.
For SAMs especially, simultaneously performing variable selection over
all archetypes is key to identifying which environmental variables are
important in structuring the species community. On the other hand,
given the number of candidate models is considerably larger than in the
standard GLM context, methods which require fitting all possible
models, for example, information criteria, are impractical.

In this article, we consider using penalized likelihood methods for
variable selection in FMR models and SAMs. Since each covariate in an
FMR model is represented by a group of coefficients, one for each
component (and whose true value may be zero), we propose two penalties
which exploit this grouped structure. This leads to an attractive form
of shrinkage that allows a covariate to be removed from all components
simultaneously. The first penalty proposed is a modification of the
group LASSO [\citet{yuan06}] to FMR models, called MIXGL2 (since it is
an $\ell_2$-norm penalty), which is applied across components on a per
covariate basis. The second penalty, called MIXGL1, is based on the
square root of the $\ell_1$-norm and allows the component densities to
depend on different sets of covariates. In a diverging number of
covariates settings (i.e., the number of parameters grows a slower rate
than sample size), we demonstrate that MIXGL1 and MIXGL2 each possess a
specific form of variable selection
consistency. Furthermore, simulation studies show MIXGL2 and MIXGL1
outperform other penalties which do not take into account the grouped
structure of the covariates, both in variable selection and prediction.

We apply both penalties to construct multi-species distribution models
for the aforementioned Great Barrier Reef data set. This data set was
collected as part of a larger biodiversity project aimed at identifying
the key environmental variables important in structuring seabed
biodiversity, as well as predicting future distributions of species
communities along the Great Barrier Reef [\citet{pitcher2007}]. It
consists of presence--absence data of 200 species collected at 1189
sites, along with 13 environmental covariates. Compared to the results
from unpenalized SAMs [\citet{dunstan11,hui13}], the penalized versions
offer two advantages: (1) clearer insight is gained into species
co-occurrence, as the penalties provide an automated way of identifying
the variables informing each archetypal response; (2) the inclusion of
penalties smooths the likelihood and leads to a more stable estimation
procedure for SAMs.

We conclude this introduction by reviewing previous literature on
penalized likelihood methods for FMR models. \citet{khalili07} were the
first to extend the LASSO [\citet{tibshirani96}] and SCAD [\citet{fan01}]
penalties to FMR models by applying these penalties on a per component
basis, that is, each component has its own tuning parameter. For FMR
models with a diverging number of covariates in each component, \citet
{khalili13} proposed elastic-net type penalties [i.e., a linear
combination of a sparsity-inducing and a ridge penalty, \citet{zou05}],
which were also applied on a per component basis. For mixtures of
linear regression in particular, \citet{stadler10} applied the adaptive
LASSO [\citet{zou06}] with one tuning parameter for the entire model. To
date, no penalty has been proposed which exploits the grouped structure
of the covariates in FMR models, something we investigate in this article.

\section{Finite mixture of regression models} \label{secfmrs}
Consider a sample of observations $\{(\mathbf{x}_i, y_i); i = 1, \ldots,n\}
$, where $y_i$ is a univariate, independent and identically distributed
response and $\mathbf{x}_i$ is a $p \times1$ vector of covariates. We
allow $p$ to grow polynomially with sample size, that is, $p^{\nu}/n
\rightarrow0$ for some $\nu> 1$. The precise value of $\nu$ is
specified later on in Section~\ref{secasympprop}. All covariates are
assumed to have been standardized to mean zero and variance one. For an
FMR model with $K$ components, the conditional density function for
observation $i$ is given as follows:
\begin{equation}
\label{eqnfmrdefine}
h(y_i;\mathbf{x}_i,\bolds{\Psi}) = \sum
_{k=1}^K \pi_k
f(y_i; \mathbf{x}_i,\mu _{ik},
\phi_k);\qquad  g(\mu_{ik}) = \beta_{0k} + \sum
_{l=1}^{p} x_{il}\beta_{kl},
\end{equation}
where $\bolds{\pi} = (\pi_1,\ldots,\pi_K)$ denotes the mixing proportions
satisfying $\pi_k > 0$,  $\sum_{k=1}^K \pi_k = 1$ and $f(y;\mathbf
{x},\mu_k,\phi_k)$ is the $k{\mbox{th}}$ component density assumed to
come from the exponential family with mean $\mu_k$ and dispersion
parameter $\phi_k$. For observation $i$, the mean conditional on
belonging to the $k{\mbox{th}}$ component, $\mu_{ik}$, is regressed
against covariates $\mathbf{x}_i$ using a GLM with link function $g(\cdot)$
and coefficients $\{\beta_{kl}; l = 1,\ldots,p\}$.

Let $\bolds{\beta}_l = (\beta_{1l}, \ldots, \beta_{Kl})$ be the vector of
coefficients corresponding to covariate $l$. Notice the coefficients
are concatenated on a \emph{per covariate} basis---this reflects the
grouped structure of the covariates, that is, each covariate is
represented by $K$ coefficients, one for each component and whose true
value may be zero. Finally, let $\bolds{\beta} = (\bolds{\beta}_1, \ldots, \bolds
{\beta}_{p})$ be the $K \times p$ matrix of regression coefficients,
and $\bolds{\Psi} = (\bolds{\beta}_1, \ldots, \bolds{\beta}_{p}, \beta_{01},
\ldots, \beta_{0K}, \bolds{\phi}, \bolds{\pi})$ denote all the parameters in
the FMR model, where $\bolds{\phi} = (\phi_1, \ldots, \phi_K)$.

In this article, it is assumed the parameters in the FMR model in
equation~(\ref{eqnfmrdefine}) are generically identifiable up to a
permutation of the component labels [see condition~(A1) in the
Supplementary Material, \citet{hui15grass}]. Furthermore, we develop our
asymptotic theory assuming the number of components $K$ is known
[analogous to \citet{stadler10,khalili13}], although for our application
with SAMs we propose a BIC-type criterion to select the number of
archetypes. General discussions regarding parameter identifiability for
mixture models can be found in \citet{mclachlan00} and \citet
{schnatter06}, with the specific case of generic identifiability
discussed in \citet{follmann91}, \citet{henning00} and \citet{grun08},
among others.

\section{New penalties for variable selection} \label{secnewpens}
To exploit the grouped structure of covariates in FMR models and SAMs,
we consider penalized likelihood methods using penalties which are
applied across components on a per covariate basis,
\[
\ell^{\mathrm{pen}}_n(\bolds{\Psi}) = \ell_n(\bolds{
\Psi}) - n\lambda\sum_{l=1}^{p} \mathcal{P}(
\bolds{\beta}_l),
\]
where $\ell_n(\bolds{\Psi}) = \sum_{i=1}^n \log (\sum_{k=1}^K \pi_k f(y_i; \mathbf{x}_i,\mu_{ik},\phi_k) )$ is the observed
log-like\-lihood and $\mathcal{P}(\bolds{\beta}_l)$ denotes a penalty
function which is nonnegative and satisfies $\mathcal{P}(\mathbf{0}) = 0$.
Let $\tilde{\bolds{\beta}} = (\tilde{\bolds{\beta}}_1, \ldots, \tilde{\bolds
{\beta}}_{p})$ denote the unpenalized maximum likelihood estimates of
$\bolds{\beta}$. We propose two penalty forms for $\mathcal{P}(\bolds{\beta
}_l)$. The first is a modification of the group LASSO [\citet{yuan06}]
for FMR models.

\begin{defin} \label{defgplassofmr}
For the FMR model defined in equation~(\ref{eqnfmrdefine}), the MIXGL2
estimates are given by maximizing the penalized log-likelihood function
\begin{eqnarray*}
&& \ell^{\mathrm{pen}}_n(\bolds{\Psi}) = \ell_n(\bolds{
\Psi}) - n\lambda\sum_{l=1}^{p}
\tilde{w}_l \sqrt{\sum
_{k=1}^K \beta^2_{kl}},
\end{eqnarray*}
where $\tilde{w}_l =  (\sum_{k=1}^K \tilde{\beta
}^2_{kl} )^{-\gamma/2}$ and $\gamma> 0$.
\end{defin}

MIXGL2 possesses the group sparsity property, that is, it is
nondifferentiable when $\beta_{1l} = \cdots= \beta_{Kl} = 0$ for
covariate $l$. This is an attractive property to have for variable
selection in FMR models, as it encourages a covariate to be removed
from all components simultaneously. Such a form of sparsity in the
solution is useful for multi-species distribution modeling, as often
there are numerous environmental covariates which are completely
uninformative for all archetypes (a covariate is defined as completely
uninformative if all its coefficients are truly zero). MIXGL2 is useful
for screening these covariates out, potentially as a first stage in
variable selection for SAMs.

The second penalty we propose is based on the square root of a weighted
$\ell_1$-norm.

\begin{defin} \label{defmcapfmr}
For the FMR model defined in equation~(\ref{eqnfmrdefine}), the MIXGL1
estimates are given by maximizing the penalized log-likelihood function
\begin{eqnarray*}
&& \ell^{\mathrm{pen}}_n(\bolds{\Psi}) = \ell_n(\bolds{
\Psi}) - n\lambda\sum_{l=1}^{p} \sqrt
{\sum_{k=1}^K
\tilde{w}_{kl} |\beta_{kl}|},
\end{eqnarray*}
where $\tilde{w}_{kl} = |\tilde{\beta}_{kl}|^{-\gamma}$ and $\gamma> 0$.
\end{defin}

MIXGL1 not only possesses the group sparsity property like MIXGL2, it
also possesses individual coefficient sparsity analogous to the
adaptive LASSO, that is, it is also nondifferentiable for all
individual coefficients $\beta_{kl}$. This individual sparsity allows
MIXGL1 to remove covariates from only $K' < K$ components. It is
therefore well suited to species distribution modeling---since the
archetypal responses typically depend on different sets of covariates,
MIXGL1 can accommodate for differing mean structures in each archetype.
Put another way, the set of variables that drive the co-occurrences of
one group of species are usually slightly different to those that drive
the co-occurrence of another group. The form of MIXGL1 allows for this,
while maintaining the ability to remove completely uninformative
covariates from the entire SAM. Of course, the choice of which penalty
also depends on the analysis objectives---sometimes, it is of interest
to see which covariates affect any (or all)
components, in which case MIXGL2 is more appropriate. Other times, we
may want to know how each covariate affects the archetypes in the most
compact way, in which case MIXGL1 is suitable.

Definition~\ref{defmcapfmr} is a special case of the Composite
Absolute Penalty (CAP) family of penalties [\citet{zhao09}], although
our work is the first to apply such a penalty to the FMR model context.
Both MIXGL2 and MIXGL1 incorporate data-dependent weights based on the
unpenalized estimates, $\tilde{\bolds{\beta}}$, with the severity of these
weights controlled by $\gamma$. The inclusion of weights builds on the
idea of the adaptive LASSO and allows the penalized estimates to
achieve desirable large sample properties as discussed in the next
section. Finally, unlike the penalties in \citet{khalili07} and \citet
{khalili13} which are applied on a per component basis, MIXGL1 and
MIXGL2 do not depend on the mixing proportions. When penalization
occurs on a per component basis, having penalties which are a function
of $\bolds{\pi}$ makes sense since it relates the severity of penalization
to the ``effective sample size'' of each component. For penalization
across components on a per covariate basis, specifically,
the MIXGL1 and MIXGL2 penalties, the need to incorporate mixing proportions is less obvious.

\subsection{Asymptotic properties} \label{secasympprop}
In this section, we consider the large sample behavior of the MIXGL2
and MIXGL1 estimators. As mentioned previously, we assume the number of
components $K$ is fixed and known {a priori}, but the number of
covariates in each component grows with sample size $n$. We shall use
$p_n$, as well as a subscript $n$ in other quantities, for example,
$\lambda_n$ and $\bolds{\beta}_n$, to reflect this. Let $\bolds{\Psi}_n^0 =
(\bolds{\beta}^0_{n,1}, \ldots, \bolds{\beta}^0_{n,p_n}, \beta^0_{n,01},
\ldots, \beta^0_{n,0K}, \bolds{\phi}_n^0, \bolds{\pi}_n^0)$ be parameter
values corresponding to the true model, which is assumed to be
identifiable. We can partition all the regression coefficients in the
true model as follows:

\begin{defin}
The regression coefficients in the true model, $(\bolds{\beta}^0_{n,1},
\ldots,\break \bolds{\beta}^0_{n,p_n})$ can be partitioned into the following
mutually exclusive sets:
\begin{itemize}
\item$\mathbf{\mathcal{A}}_n = \{(k,l)\dvtx  \beta^0_{n,kl} \ne0\}$ denotes
the set of truly nonzero coefficients.
\item$\mathbf{\mathcal{B}}_n = \{(k,l)\dvtx  \beta^0_{n,kl} = 0, \|\bolds{\beta
}^0_{j}\|_2 \ne0\}$ denotes the set of zero coefficients belonging to
partly uninformative covariates.
\item$\mathbf{\mathcal{C}}_n = \{(k,l)\dvtx  \beta^0_{n,kl} = 0, \|\bolds{\beta
}^0_{j}\|_2 = 0\}$ is the set of zero coefficients belonging to
completely uninformative covariates.
\end{itemize}
\end{defin}

As formalized below, the group sparsity property of MIXGL2 allows it to
asymptotically set all coefficients belonging to set $\mathbf{\mathcal{C}}_n$ to zero, while the combined group and individual coefficient
sparsity property of MIXGL1 allows it to asymptotically set all
coefficients belonging to sets $\mathbf{\mathcal{B}}_n$ and $\mathbf{\mathcal{C}}_n$ to zero.

For both penalties, assume the following regularity conditions are
satisfied:
\begin{longlist}[(A$^\prime$)]
\item[(A)]  $\lambda_n a_n = o_p(n^{-1/2})$,
\item[(A$^\prime$)] $\lambda_n a_n
= o_p(n^{-1/2}p^{-1}_n)$,\vspace*{1.5pt}
\item[(B)] $p_n^2/(\lambda_n^2 b_n) = o_p(n)$,\vspace*{1.5pt}
\item[(C)] $p_n^4/n \rightarrow0$,\vspace*{1.5pt}
\item[(C$^\prime$)] $p_n^5/n
\rightarrow0$,
\end{longlist}
where for the MIXGL2 penalty, $a_n = \max\{\tilde{w}_{n,l}; l \in
\mathbf{\mathcal{A}}_n \}$ and $b_n = \min\{\tilde{w}^2_{n,l};\break l \in
\mathbf{\mathcal{C}}_n \}$, and, analogously for the MIXGL1 penalty, $a_n =
\max\{\tilde{w}_{kl}; (k,l) \in\mathbf{\mathcal{A}}_n \}$ and $b_n = \min
\{\tilde{w}_{kl}; (k,l) \in\mathbf{\mathcal{B}}_n \cup  \mathbf{\mathcal{C}}_n\}$. Conditions (A) and (A$^\prime$) ensure the existence of
penalized likelihood estimates which are asymptotically unbiased, while
condition (B) ensures an appropriate degree of shrinkage. The rate of
growth of the number of covariates in conditions (C) and (C$^\prime$)
is the same as \citet{khalili13}, and we believe it to be appropriate
in many applications of species distribution modeling in ecology, that
is, the number of environmental variables recorded is usually small
compared to the number of sites visited.

We\vspace*{-2pt} first consider the asymptotic behavior of the MIXGL2 estimator:
\begin{thm}[(Oracle property---MIXGL2)]
\label{thmgplassoconsistency}
Assume conditions \textup{(A)}--\textup{(C)} hold. Then there exists a local maximizer
$\hat{\bolds{\Psi}}_n$ of $\ell^{\mathrm{pen}}_n(\bolds{\Psi}_n)$ in Definition~\ref{defgplassofmr} which satisfies the\vspace*{-2pt} following:
\begin{itemize}
\item Estimation consistency: $\|\hat{\bolds{\Psi}}_n - \bolds{\Psi}_n^0\| =
O_p(\sqrt{p_n/n})$.
\item Covariate selection consistency: $P(\hat{\bolds{\beta}}_{n,\mathbf{\mathcal{C}}_n} = \mathbf{0}) \rightarrow1$.
\item Asymptotic normality: If conditions (\textup{A}$^\prime$) and (\textup{C}$^\prime$)
are also satisfied,\vspace*{-3pt} then
\begin{eqnarray*}
&& \sqrt{n} \bolds{\Gamma}_n \mathcal{I}_n\bigl(\bolds{
\Psi}^0_{n,\mathbf{\mathcal{C}}^c_n}\bigr)^{1/2} \bigl(\hat{\bolds{
\Psi}}_{n,\mathbf{\mathcal{C}}^c_n} - \bolds{\Psi }^0_{n,\mathbf{\mathcal{C}}^c_n}\bigr) \mathop{
\rightarrow}^{d} N(\mathbf{0}, \mathbf{G}),
\end{eqnarray*}
\end{itemize}
where $\hat{\bolds{\Psi}}_{n,\mathbf{\mathcal{C}}^c_n} = (\hat{\bolds{\beta
}}_{n,\mathbf{\mathcal{C}}^c_n}, \hat{\beta}_{n,01}, \ldots, \hat{\beta
}_{n,0K}, \hat{\bolds{\phi}}_n, \hat{\bolds{\pi}}_n)$, $\bolds{\Gamma}_n$ is a
$q \times|\mathbf{\mathcal{C}}^c_n|$ matrix such that $\bolds{\Gamma}_n \bolds
{\Gamma}_n' \mathop{\rightarrow}^{p} \mathbf{G}$, and $\mathcal{I}_n(\bolds{\Psi
}^0_{n,\mathbf{\mathcal{C}}^c_n})$ is the Fisher information matrix knowing
$\mathbf{\mathcal{C}}^c_n$.
\end{thm}

All proofs have been relegated to the Supplementary Material [\citet
{hui15grass}]. Theorem~\ref{thmgplassoconsistency} states MIXGL2 is
\emph{covariate selection consistent}, that is, it will asymptotically
remove completely uninformative covariates from the FMR model. On the
other hand, if the true model contains partly uninformative covariates
($\mathbf{\mathcal{B}}_n \ne\varnothing$), MIXGL2 will in the large sample
limit retain these covariates in all components. This makes sense
because MIXGL2 does not possess individual coefficient sparsity. By
contrast, if we consider the asymptotic behavior of the MIXGL1
estimator, then we have the following:

\begin{thm}[(Oracle property---MIXGL1)] \label{thmmcaporacle}
Assume conditions \textup{(A)}--\textup{(C)} hold. Then there exists a local maximizer
$\hat{\bolds{\Psi}}_n$ of $\ell^{\mathrm{pen}}_n(\bolds{\Psi}_n)$ in
Definition~\ref{defmcapfmr} which satisfies the following:
\begin{itemize}
\item Estimation consistency: $\|\hat{\bolds{\Psi}}_n - \bolds{\Psi}_n^0\| =
O_p(\sqrt{p_n/n})$.
\item Coefficient selection consistency: $P(\hat{\bolds{\beta}}_{n,\mathbf{\mathcal{B}}_n \cup  \mathbf{\mathcal{C}}_n} = \mathbf{0}) \rightarrow1$.
\item Asymptotic normality: If conditions (\textup{A}$^\prime$) and (\textup{C}$^\prime$)
are also satisfied,\vspace*{-3pt} then
\begin{eqnarray*}
&& \sqrt{n} \bolds{\Gamma}_n \mathcal{I}_n\bigl(\bolds{
\Psi}^0_{n,\mathbf{\mathcal{A}}_n}\bigr)^{1/2} \bigl(\hat{\bolds{
\Psi}}_{n,\mathbf{\mathcal{A}}_n} - \bolds{\Psi }^0_{n,\mathbf{\mathcal{A}}_n}\bigr) \mathop{
\rightarrow}^{d} N(\mathbf{0}, \mathbf{G}),
\end{eqnarray*}
\end{itemize}
where $\hat{\bolds{\Psi}}_{n,\mathbf{\mathcal{A}}_n} = (\hat{\bolds{\beta
}}_{n,\mathbf{\mathcal{A}}_n}, \hat{\beta}_{n,01}, \ldots, \hat{\beta
}_{n,0K}, \hat{\bolds{\phi}}_n, \hat{\bolds{\pi}}_n)$, $\|\cdot\|$ denotes
the $\ell_2$-norm, $\bolds{\Gamma}_n$ is a $q \times|\mathbf{\mathcal{A}}_n|$
matrix such that $\bolds{\Gamma}_n \bolds{\Gamma}_n' \mathop{\rightarrow}^{p} \mathbf{G}$,
and $\mathcal{I}_n(\bolds{\Psi}^0_{n,\mathbf{\mathcal{A}}_n})$ is the Fisher
information matrix knowing\vspace*{-2pt} $\mathbf{\mathcal{A}}_n$.
\end{thm}

Theorem~\ref{thmmcaporacle} states MIXGL1 is \emph{coefficient
selection consistent}, that is, it will asymptotically remove
completely uninformative covariates \emph{and} zero coefficients
belonging to partly uninformative covariates from the FMR model. This
is a stronger form of selection consistency compared to MIXGL2, and is
a desirable property in terms of identifying the truly important
covariates facilitating species co-occurrence.

To compute the MIXGL2 and MIXGL1 estimates, we use an estimation
procedure combining the Expectation Maximization [EM, \citet
{dempster77}] algorithm with a local quadratic approximation [LQA, \citet
{fan01}]. Details regarding this estimation procedure, and a proof that
it possesses the desired ascent property, may be found in the
Supplementary Material [\citet{hui15grass}]. To select the tuning
parameters $(\lambda_n, \gamma)$, we use a BIC-type criterion [see \citet
{zhang10} for the use of information criteria in selecting tuning parameters],\vspace*{-5pt}
\begin{eqnarray*}
&& \mbox{BIC}_{\lambda_n,\gamma} = -2\ell_n(\hat{\bolds{
\Psi}}_n) + \log(n) \operatorname{dim}(\hat{\bolds{\beta}}_n),
\end{eqnarray*}
where $\dim(\hat{\bolds{\beta}}_n)$ denotes the number of nonzero
estimates in $\hat{\bolds{\beta}}_n$. Note $\hat{\bolds{\Psi}}_n$ and $\dim
(\hat{\bolds{\beta}}_n)$ both depend on $\lambda_n$ and $\gamma$, for
example, as $\lambda_n$ increases, more values of $\hat{\bolds{\beta}}_n$
are shrunk to zero. For our simulations and applications, we select
from $\gamma\in\{0.5,1,2\}$.

\section{Application} \label{secapplication}
We apply MIXGL1 and MIXGL2 to Species Archetype Models and the Great
Barrier Reef data set introduced in Section~\ref{secintro}. To recap,
the data set consists of presence--absence data of 200 species at $1189$
sites, along with 13 environmental covariates. Five of these covariates
were descriptors of physical habitat: depth (BATHY), bottom stress
(BSTRESS), percent gravel (GRAVEL), percent mud (MUD), percent
carbonate (CARBON), and the other eight covariates were oceanographic
measures: mean and intra-annual standard deviation of temperature
(T.AV, T.SD), mean and intra-annual standard deviation of oxygen
concentration (O2.AV, O2.SD), mean and intra-annual standard deviation
of salinity (S.AV, S.SD), mean and intra-annual standard deviation of
K490 (K490.AV,\break K490.SD). K490 is measure of turbidity based on light
penetration and is related to the presence of light scattering
particles in the water.

To model the 200 species separately using a GLM (say) would be a
difficult task, especially since 164 out of 200 species are present at
less than 5\% of the sites. Such sparsity in the response (in the sense
that most species are rarely observed) is characteristic of
multi-species data and motivates methods such as SAMs which are able to
borrow strength across species. On the other hand, while the
applications of SAMs in \citet{dunstan11} and \citet{dunstan13} were
mainly for illustrative purposes, the goal in this article is to
perform (consistent) variable selection in order to identify the key
drivers of species co-location.

Let $\mathbf{Y}$ denote the multi-species response matrix collected $i =
1,\ldots,n$ sites for $j = 1,\ldots,s$ species, with $[\mathbf{Y}]_{ij} =
1$ if species $j$ was found at site $i$ and 0 otherwise. For the Great
Barrier Reef data set, $n = 1189$ and $s = 200$. Let $\mathbf{x}_i$ be the
vector of environmental covariates at site $i$. SAMs are an extension
of FMR models in (\ref{eqnfmrdefine}) to the case of product Bernoulli
component\vspace*{-3pt} densities:
\begin{eqnarray}
 h(\mathbf{y}_j; \mathbf{x}_i,\bolds{
\Psi}) &=& \sum_{k=1}^K \pi_k
\Biggl(\prod_{i=1}^n \mu_{ijk}^{y_{ij}}(1-
\mu_{ijk})^{1-y_{ij}} \Biggr); \nonumber
\\[-10pt]
\label{eqnsam}
\\[-10pt]
\nonumber
 \operatorname{logit}(\mu_{ijk}) &=&
\beta_{0j} + \sum_{l=1}^p
x_{il}\beta_{kl}.
\end{eqnarray}
Note in (\ref{eqnsam}), the intercepts $\beta_{0j}$ are
species-specific instead of component-specific. As discussed in \citet
{dunstan13}, this is so that species are clustered solely on the shape
of their environmental responses and not as well on the prevalence.

For each of the 13 covariates, we fitted linear and quadratic terms,
resulting in 26 terms available for selection in each archetype. We
used MIXGL2 and MIXGL1 to perform variable selection, with the tuning
parameters chosen using $\mbox{BIC}_{\lambda_n,\gamma}$ defined in
Section~\ref{secasympprop}, and $\log(n)$ replaced by $\log(s)$ as the
model complexity penalty. The reason for this replacement is because
the fundamental observational unit in SAMs is a species instead of a
site [see \citet{dunstan11} and also the Discussion in Section~\ref{secdiscussion}]. For a fixed $K$, the combined EM plus LQA algorithm
used for fitting penalized FMR models can be straightforwardly extended
to the case of SAMs [see the Supplementary Material for details, \citet
{hui15grass}]. To select the number of archetypes, we considered a
candidate range $K = [1,20]$ and used another BIC-type criterion [see
Sections~6.8--6.9, \citet{mclachlan00}, on using information criteria for
selecting\vspace*{-2pt} $K$],
\begin{eqnarray*}
&& \mbox{BIC}_{K} = -2\ell_n\bigl(\hat{\bolds{
\Psi}}(K)_n\bigr) + \log(s) \dim\bigl(\hat{\bolds {
\Psi}}(K)_n\bigr),
\end{eqnarray*}
where $\dim(\hat{\bolds{\Psi}}(K))_n$ denotes the number of nonzero
parameters in the SAM with $K$ archetypes. The model selection
procedure thus proceeds as follows: for each candidate $K$, the best
penalized SAM is selected using $\mbox{BIC}_{\lambda_n,\gamma}$.
Afterward, the final model is selected from these ``best penalized
SAMs'' using $\mbox{BIC}_{K}$.

\begin{figure}

\includegraphics{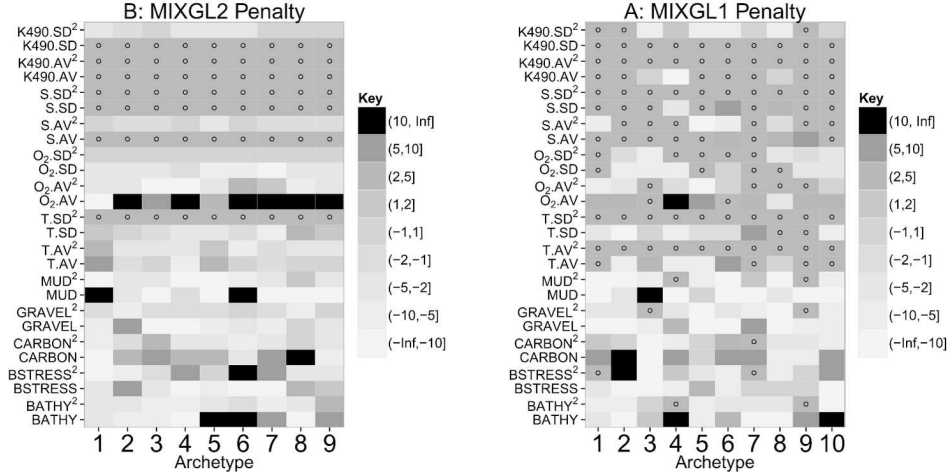}
\vspace*{-3pt}
\caption{Map of coefficients for penalized SAMs with MIXGL2 (left) and
MIXGL1 penalties (right). Empty circles indicate coefficients shrunk to
exactly 0. Based on $\mbox{BIC}_{K}$, the former chose $K = 9$
archetypes while the latter chose $K = 10$ archetypes.}
\label{figcoefmap}
\end{figure}

Figure~\ref{figcoefmap} shows a plot of coefficients estimates for the
two penalized SAMs, with exact values of the coefficients provided in
the Supplementary Material [\citet{hui15grass}]. Using $\mbox{BIC}_{K}$,
MIXGL2 and MIXGL1 produced SAMs with $K=9$ and 10 archetypes,
respectively [note \citet{dunstan11} chose $K=11$ archetypes when
applying unpenalized SAMs to the same data set]. Compared with the
unpenalized SAM in \citet{dunstan11}, penalized variable selection
offers much more insight into the drivers of seabed biodiversity. Both
models indicated the five physical habitat covariates were informative
for almost all archetypes (Figure~\ref{figcoefmap}---bottom 10 rows in
both plots), meaning these variables are important in structuring the
entire marine species assemblage. In contrast, MIXGL1 and MIXGL2 deemed
several of the oceanographic measures to be either completely or close
to completely uninformative, for example, the quadratic terms T.SD$^2$
and S.SD$^2$. It is also in these
oceanographic covariates where the archetypal responses were largely
differentiated for the MIXGL1 model. For example, archetype 2 was the
only archetype where the four terms corresponding to mean and
intra-annual standard deviation of oxygen were informative. That is,
the co-occurrence of species classified to this archetype could be
partly attributed to their shared environmental response to oxygen.
Also, archetypes 2 to 6 all had coefficients for mean annual
temperature which were substantially different from zero. However,
species in archetypes 3 and 6 tend to occupy regions of higher
temperatures, while species classified to archetype 2, 4 and 5 tend to
occur in relatively cooler regions.

For prediction purposes, maps were constructed for each archetype
showing how the probability of presence on the linear predictor scale
varies spatially along the entire Great Barrier Reef. That is, for
archetype $k = 1,\ldots,K$, these were constructed using the linear predictor,
\begin{eqnarray*}
&& \hat{\eta}_{ik} = \frac{\sum_{j=1}^s \hat{\tau}_{jk} \hat{\beta
}_{0j}}{\sum_{j=1}^s \hat{\tau}_{jk}} + \sum
_{l=1}^p x_{il}\hat{\beta}_{kl},
\end{eqnarray*}
which were then mapped across all sites in the Barrier Reef region. To
clarify, we chose to map the linear predictors $\hat{\eta}_{ik}$
directly rather than convert them to probabilities, as this generally
makes it easier to identify any differences between the archetypes.
Note the intercept used in the predicted maps is a weighted average of
all species-specific intercepts, with weights proportional to the
posterior probabilities of belonging to archetype $k$. These maps are
provided in the Supplementary Material [\citet{hui15grass}]. Note that
maps could also be constructed for each species, although for
managerial purposes maps constructed on a per archetype basis tend to
be more useful, since managing an archetype is equivalent to
simultaneously managing all the species classified into the archetype
[\citet{dunstan11}].

Both MIXGL2 and MIXGL1 produced some similar maps, for example,
archetype 4 in both models exhibited a high probability of presence
from the central to the southeast region of the reef toward the Coral
sea, with the probability decreasing sharply toward land. Also, species
in archetype 9 for both models tend to be found with relatively high
probability toward the southeast Queensland coast. Given the differing
properties of MIXGL2 and MIXGL1, however, it was not surprising to also
observe some notable differences between the two sets of maps, for
example, archetype 2 for the MIXGL2 fit displayed relatively high
probabilities of presence in a small region on the southeast region of
the Barrier Reef, where percent gravel was rather high. However, no
corresponding archetype was observed for the MIXGL1 fit---while
archetypes 4
and 7 also had positive linear terms for percent gravel (Figure~\ref{figcoefmap}---right), the environmental response of species in these
two components tends to be driven more by other covariates. This
suggests incorporating a penalty such as MIXGL1, which allows different
mean structures in each archetype and leads to more precise
differentiation of the sources of species co-occurrences.

\section{Simulation study} \label{secsims}
A small simulation was performed to assess the finite sample
performance of the MIXGL1 and MIXGL2 in FMR models in equation~(\ref
{eqnfmrdefine}). The number of covariates in each component was
determined as $p_n = \lceil4n^{1/4} \rceil- 5$, where $\lceil\cdot
\rceil$ is the ceiling function [see also \citet{khalili13}]. Covariates
$\{\mathbf{x}_i; i = 1,\ldots,n\}$ were generated from a standard
multivariate Gaussian distribution with correlation structure $\operatorname{Cor}(x_{ir}, x_{is}) = 0.5^{|r-s|}$. Responses were then simulated
from a $K=2$ binomial FMR model with trial size 10, using the four
models below:
\[
\begin{array} {cccccccccccc} & (\beta_{01},\bolds{
\beta}_1) &= (&\!\!1,&0.7,&2,&-2,&1.5,&0,&0,&0,&\ldots),
\\
\mbox{Model I:} & (\beta_{02},\bolds{\beta}_2) &=
(&\!\!{-}0.5,&2,&0,&0,&0,&1,&-2,&0.5,&\ldots),
\\
\mbox{Model II:} & (\beta_{02},\bolds{\beta}_2) &=
(&\!\!{-}0.5,&2,&0,&0,&1,&-2,&0.5,&0,&\ldots),
\\
\mbox{Model III:} & (\beta_{02},\bolds{\beta}_2) &=
(&\!\!{-}0.5,&2,&0,&1,&-2,&0.5,&0,&0,&\ldots),
\\
\mbox{Model IV:} & (\beta_{02},\bolds{\beta}_2) &=
(&\!\!{-}0.5,&2,&1,&-2,&0.5,&0,&0,&0,&\ldots), \end{array} %
\]
where ``$\ldots$'' indicates extra zeros. The models are designed such
that, as we move from models I to IV, the number of partly
uninformative covariates decreases. We considered combinations of $\pi
_1 = 0.5,0.7$ and $n = 100,200,400$, with the latter corresponding to
$p_n = 7,9,12$ covariates (excluding intercept) in each component,
respectively. 500 data sets were generated for each combination. We
assumed $K=2$ was known in advance.

\begin{table}[b]
\tabcolsep=0pt
\caption{Mean sensitivity/specificity for various sample sizes and $\pi
_1 = 0.5$. Transitioning from models \textup{I} to \textup{IV}, the proportion of
completely uninformative covariates increases} \label{tabsensispeci}
\begin{tabular*}{\tablewidth}{@{\extracolsep{\fill}}lcccccc@{}}
\hline
& & \multicolumn{5}{c@{}}{\textbf{Sensitivity/Specificity}} \\[-4pt]
&&\multicolumn{5}{l@{}}{\hrulefill}\\
$\bolds{n}$ & \textbf{Model} & \textbf{MIXGL1} & \textbf{MIXGL2} & \textbf{ADL} & \textbf{MIXLASSO-}$\bolds{\ell_2}$ &
\textbf{MIXSCAD-}$\bolds{\ell_2}$ \\
\hline
100  & \phantom{II}I & $0.962\mbox{/}0.948$ & $0.947\mbox{/}0.040$ & $0.958\mbox{/}0.897$ &
$0.991\mbox{/}0.170$ & $0.978\mbox{/}0.690$ \\
& \phantom{I}II & $0.962\mbox{/}0.962$ & $0.959\mbox{/}0.373$ & $0.956\mbox{/}0.855$ & $0.992\mbox{/}0.128$ &
$0.975\mbox{/}0.683$ \\
& III & $0.966\mbox{/}0.972$ & $0.957\mbox{/}0.747$ & $0.950\mbox{/}0.832$ & $0.989\mbox{/}0.160$ &
$0.970\mbox{/}0.647$ \\
& IV & $0.957\mbox{/}0.980$ & $1\mbox{/}0.970$ & $0.965\mbox{/}0.825$ & $0.981\mbox{/}0.183$ & $0.970\mbox{/}0.635$
\\[3pt]
200 & \phantom{II}I & $0.983\mbox{/}0.986$ & $0.960\mbox{/}0.320$ & $0.994\mbox{/}0.948$ &
$0.998\mbox{/}0.388$ & $0.996\mbox{/}0.826$ \\
& \phantom{I}II & $0.986\mbox{/}0.983$ & $0.954\mbox{/}0.651$ & $0.987\mbox{/}0.935$ & $0.999\mbox{/}0.430$ &
$0.993\mbox{/}0.819$ \\
& III & $0.985\mbox{/}0.985$ & $0.956\mbox{/}0.864$ & $0.994\mbox{/}0.911$ & $0.999\mbox{/}0.415$ &
$0.995\mbox{/}0.803$ \\
& IV & $0.985\mbox{/}0.989$ & $1\mbox{/}1$ & $0.987\mbox{/}0.908$ & $0.995\mbox{/}0.392$ & $0.991\mbox{/}0.782$ \\[3pt]
400 & \phantom{II}I & $0.998\mbox{/}0.992$ & $0.979\mbox{/}0.636$ & $0.999\mbox{/}0.961$ &
$1\mbox{/}0.546$ & $1\mbox{/}0.918$ \\
& \phantom{I}II & $0.999\mbox{/}0.991$ & $0.978\mbox{/}0.761$ & $0.999\mbox{/}0.958$ & $1\mbox{/}0.576$ & $0.999\mbox{/}0.901$
\\
& III & $0.999\mbox{/}0.995$ & $0.981\mbox{/}0.884$ & $0.999\mbox{/}0.940$ & $1\mbox{/}0.561$ & $0.999\mbox{/}0.897$
\\
& IV & $0.999\mbox{/}0.992$ & $1\mbox{/}1$ & $0.997\mbox{/}0.942$ & $1\mbox{/}0.543$ & $0.998\mbox{/}0.895$ \\
\hline
\end{tabular*}
\end{table}

We compared MIXGL1 and MIXGL2 to three penalties proposed previously
for FMR models: adaptive LASSO [ADL, \citet{stadler10}], MIXLASSO-$\ell_2$, and MIXSCAD-$\ell_2$ [\citet{khalili13}]. The latter two penalties
are linear combinations of the ridge and LASSO (SCAD) penalty. ADL
penalizes coefficients separately, while\break MIXLASSO-$\ell_2$ and
MIXSCAD-$\ell_2$ penalize on a per component basis. Performance was
assessed using mean sensitivity (proportion of true nonzeros estimated
to be nonzero) and specificity (proportion of true zeros estimated to
be zero), and predicted log-likelihood. The latter was calculated using
an independent test data set of $n = 10{,}000$ observations, with higher
values implying better predictions. Note that to remove variation
across the different data sets, we centered the predicted
log-likelihood values by subtracting the average obtained across the
different methods in each data set. Also, to deal with the problem of
label-switching prior to calculating sensitivity and
specificity [see Section~4.9, \citet{mclachlan00}], we permuted the
estimated coefficients so as to minimize the $\ell_2$-norm between the
estimated and true coefficients. For brevity, we only present results
for $\pi_1 = 0.5$. Similar outcomes were observed for $\pi_1 = 0.7$,
and these results are provided in the Supplementary Material [\citet
{hui15grass}].


MIXGL1 performed consistently well, with sensitivity and specificity
lar\-gely unaffected by the number of completely uninformative covariates
(Table~\ref{tabsensispeci}). Compared to MIXGL1, the three methods
which did not penalize on a per-covariate basis had lower specificity
indicative of overfitting. The result makes sense given the group
sparsity property of MIXGL1 and MIXGL2---for all four models, there
was a proportion of coefficients which always belonged to completely
uninformative covariates. Therefore, penalties which can remove a
covariate from all components simultaneously were much better at
shrinking these coefficients to zero. As the number of partly
uninformative covariates decreased, the performance of MIXGL2
dramatically improved, especially in specificity. Another interesting
trend observed when Transitioning from models I to IV, in this
simulation at least, was a slight but noticeable decline in specificity
for ADL and MIXSCAD-$\ell_2$ (indicating overfitting). Finally,
MIXLASSO-$\ell_
2$ had the highest sensitivity in most models, but significantly lower
specificity than all the other methods tested, which suggested a
substantial amount of overfitting.

\begin{figure}[t]

\includegraphics{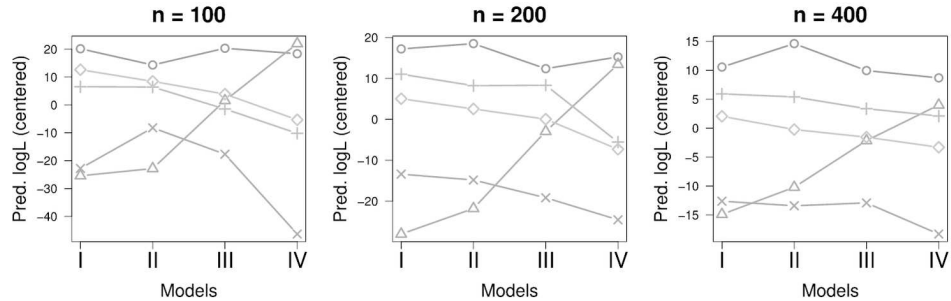}

\caption{Predicted log-likelihood (centered) as a function of a
simulation model for $\pi_1 = 0.5$. The methods shown are the
following: MIXGL1 ($\bigcirc$), MIXGL2 ($\bigtriangleup$), ADL ($+$),
MIXLASSO-$\ell_2$ ($\times$), MIXSCAD-$\ell_2$ ($\Diamond$). As we move
from models \textup{I} to \textup{IV}, the proportion of completely uninformative
covariates increases. Note the different scales on the $y$-axis for the
three figures.} \label{figpredlogl}
\end{figure}

The trends in sensitivity and specificity (Table~\ref{tabsensispeci})
were similarly observed in predictive performance. MIXGL1 predicted the
best overall, while as we move from models I to IV and decrease the
number of partly uninformative covariates, predictions from MIXGL2
improved significantly (Figure~\ref{figpredlogl}). There is also a
slight decreasing trend in predictive log-likelihood for ADL and
MIXSCAD-$\ell_2$, which appeared to coincide with a slight drop in
specificity. In the Supplementary Material [\citet{hui15grass}], we
present an additional simulation conducted with mixtures of linear
regression, $K=4$, and more covariates, with similar results to those
observed above despite the larger number of components.

\section{Discussion} \label{secdiscussion}
Species in a community exhibit significant heterogeneity in their
occurrence patterns. A major source of this heterogeneity is due to
species responses being driven by different but potentially overlapping
sets of environmental covariates. In the context of multi-species
distribution modeling using SAMs, this means we require a method of
variable selection which can consistently identify the covariates
responsible for shaping each archetypal response (and thus shaping
species co-occurrences within each archetype). In this article, we
proposed two penalties, MIXGL2 and MIXGL1, which exploit the grouped
structure of covariates in FMR models and SAMs. By penalizing across
components on a per covariate basis, these penalties can remove a
covariate simultaneously from all components of an FMR model. Both
penalties were shown to posses specific forms of selection consistency,
with simulations indicating they outperform other penalties which do
not take into account the grouped structure of covariates.
Applying penalized SAMs to the Great Barrier Reef data set offered
clearer insight into the variables structuring seabed biodiversity,
while providing a relatively simple idea of how the species assemblage
as a whole responds to the environment.

While we have demonstrated that penalized SAMs can help to unravel how
the distribution of a species community depends on environmental
covariates, imposing a penalty on the likelihood also offers
computational advantages. Given the high dimensionality ($s/n$ being a
nonnegligible ratio) and heterogeneity in environmental responses of
multi-species data sets, the likelihood for a SAM is expected to be
``bumpy'' with numerous local maxima. The estimates obtained from
applying the EM algorithm to an unpenalized SAM may therefore depend
heavily on the starting point, and may correspond to a local instead of
the (one of $K!$ equivalent) global maximum. Adding a penalty to the
likelihood can help to resolve this problem by smoothing the likelihood
surface and making the global maxima more apparent.

\begin{figure}

\includegraphics{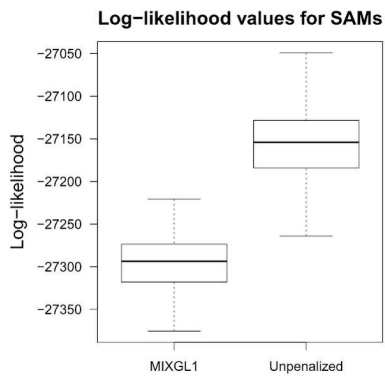}

\caption{Comparative boxplot of the log-likelihood values from 50 fits
of a penalized SAM with the MIXGL1 penalty (left) and an unpenalized
SAM (right). Both models fitted the same set of covariates and the same
number of archetypes.} \label{figproflogl}
\end{figure}

As an illustration of this, we considered two models fitted to the
Great Barrier Reef data set: (1) the MIXGL1 penalized SAM with $K=10$
and the tuning parameter fixed at it the final value used in
Section~\ref{secapplication}; (2) an unpenalized SAM, that is,
$\lambda= 0$, with $K=10$ and the 26 covariate terms included in each
archetype. Each model was fitted 50 times using the same estimation
procedure as in Section~\ref{secapplication}, each time using a random
starting point generated by simulated posterior probabilities for each
species from a Dirichlet distribution with hyperparameters all set to
1. Figure~\ref{figproflogl} shows a comparative boxplot of the
resulting log-likelihood values. Importantly, the variability of the
log-likelihood values for the penalized SAM was smaller compared to the
unpenalized SAM (ratio of variance${}={}$1.510; $F$-test $p$-value${}<{}$0.01).
The reduction in variability can be attributed to the MIXGL1 penalty
smoothing the SAM likelihood surface, removing some of the ``bumps and
small hills'' and leading to a more stable estimation algorithm.

In future work, we hope to extend penalized SAMs to multi-species
pre\-sence-only data, particularly given the commonality of such data and
recently shown equivalences between point process models and
Poisson/Logistic GLMs [\citet{warton10,fithian13}]. Whether the
consistency and oracle properties of MIXGL1 and MIXGL2 hold in this
context should be considered. Also, other penalties which exploit the
grouped structure of the covariates should be considered [e.g., the
sparse group LASSO, \citet{simon2013sparse}]. In our application to the
Barrier Reef data set, MIXGL2 and MIXGL1 did not take into account the
hierarchical structure of polynomials, for example, the linear term for
intra-annual standard deviation of K490 (K490.SD) was removed while the
quadratic term remained in the model. While this still makes sense
ecologically (i.e., given all covariates were centered, then the values
of K490.SD where species are most likely to be found was around the
average value observed in the data set), a penalty which
explicitly obeys this hierarchical principle would be preferred, such
as the fused lasso [\citet{tibshirani2005sparsity}]. Finally, the
validity of BIC or any other information criterion for choosing the
tuning parameter in MIXGL1 and MIXGL2 remains an open question. In
particular, whether the model complexity penalty for SAMs should be
modified to $\log(s)$ (as was done in this article to reflect the
fundamental observational unit being a species), remain as $\log(n)$ or
perhaps be something else [see, e.g., \citet{hui14eric}] warrants further investigation.

\section*{Acknowledgments}
Thanks to the Associate Editor, two anonymous reviewers and Bill
Venables for useful discussions.

\begin{supplement}[id=suppA]
\stitle{Supplement to ``Multi-species distribution modeling using penalized mixture of
regressions''}
\slink[doi]{10.1214/15-AOAS813SUPP} 
\sdatatype{.pdf}
\sfilename{aoas813\_supp.pdf}
\sdescription{Material includes technical proofs, details on
estimation procedure and additional simulation and application results.}
\end{supplement}


\printaddresses
\end{document}